\documentclass[a4paper,11pt]{article}

\usepackage[english]{babel}
\usepackage[dvips]{graphicx}
\usepackage{amsmath}
\usepackage{amssymb}
\usepackage{epsfig}

\begin{document}

\title{Characterization of Etched Glass Surfaces by Wave Scattering}
\author{G. R. Jafari, S. M. Mahdavi, A. Iraji zad and P. Kaghazchi \\
Department of Physics, Sharif University of Technology,\\
P.O. Box 11365-9161, Tehran, Iran
\\} \maketitle

\begin{abstract}
The roughness of glass surfaces after different stages of etching is
investigated by reflection measurements with a spectrophotometer, light scattering,
and atomic-force microscopy (in small scale), and Talysurf (in large scale).
The results suggest, there are three regimes during etching, according to their
optical reflectivity and roughness. The first and second regimes are studied by
the Kirchhoff theory and the third one is studied by the geometric theory. Also,
we compare the roughness obtained by the optical scattering to the AFM results.
\end{abstract}

%PACS numbers:  \\

\section{Introduction}

The technology of micro fabrication of glass is gaining in
importance because more and more glass substrates are currently
being used to fabricate micro electro mechanical system (MEMS) devices \cite{Won}.
 Glass has many advantages as a material for MEMS applications,
 such as good mechanical and optical properties, high electrical insulators,
 and it can be easily bonded to silicon substrates at temperatures lower
 than for fusion bonding \cite{Melvin}. Also micro and nano-structuring
 of glass surfaces is important for the production of many components and
 systems such as gratings, diffractive optical elements, planar wave guide devices,
micro-fluidic channels and substrates for (bio) chemical
lab-one-chip applications \cite{Esashi, Pierrat}, although wet etching is well
developed for some of these applications \cite{Knotter, Spierings}. There are different
ways to enhance the efficiency of some optical devices such as
semiconductors lasers, solar cells etc. One of possible ways is
light is allowed to have more reaction with material in which is
propagated. Consequently the surface boundaries, need to be
roughened. In planar wave guides on the other hand the surface
boundaries have to be as smooth as possible to have good light
confinement \cite{Schuitema, Glebov}. To get reliable surface roughness, it has to be
employed non-expensive and nondestructive method to measure their
roughness. In recent years, AFM has become
available tool for studying microstructure changes in material
science. This technique enables us to measure and describe the shape
material surface with minimal sample preparation \cite{Jandt1, Silikas, Jandt2}. Although
this is a precise technique, it is an expensive and difficult
method. Also we may only investigate very small surfaces by this
technique. Study of wave scattering from self-affine (fractal)
surfaces has become very active; see for example references [12-19].
%\cite{Jaggard, Shepard, McSharry, Lin, Chen, Sheppard, Sanchez, Zhao}.
Because scattering of light from random rough surfaces is a subject
of great interest, both from a theoretical point of view and for
applications, a large number of papers have been devoted to the
subject. The measurement of light scattering at rough surfaces and
its relationship with the statistical parameters and functions
describe the surface roughness and correlation length. This
technique enables us to measure roughness of large area.
%%%%%%%%%%%%%%%%%%%%%%%%%%%%%%%%%%%%%%%%%%%%%%%%%%%%%%%%%%%%%%%%%%%%%%%%%%
\begin{center}
\begingroup
\begin{figure}
(a)\hspace{4mm}\includegraphics[width=5.3cm,height=4.9cm,angle=0]{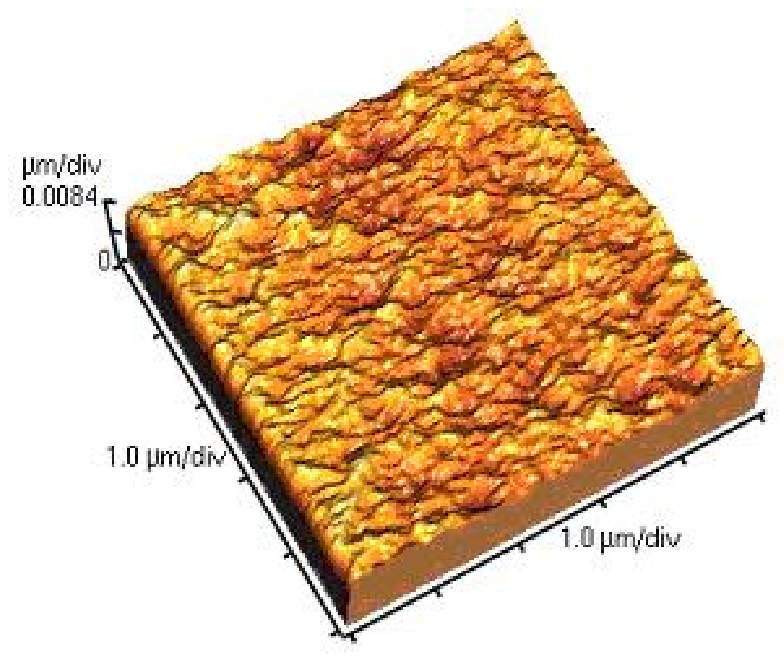}
\includegraphics[width=5.4cm,height=5cm,angle=0]{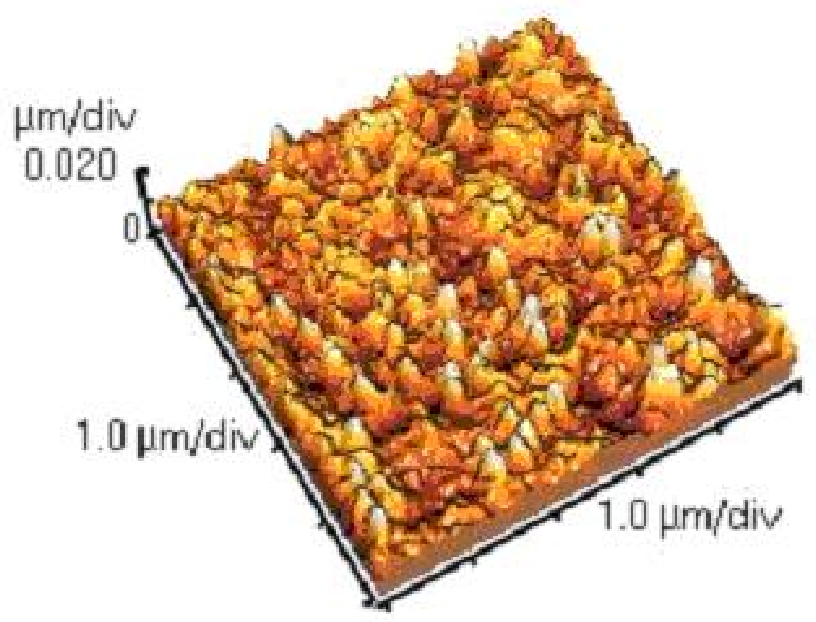}\hspace{4mm}(b)\\
\vspace{3mm}
(c)\hspace{4mm}\includegraphics[width=5.3cm,height=4.9cm,angle=0]{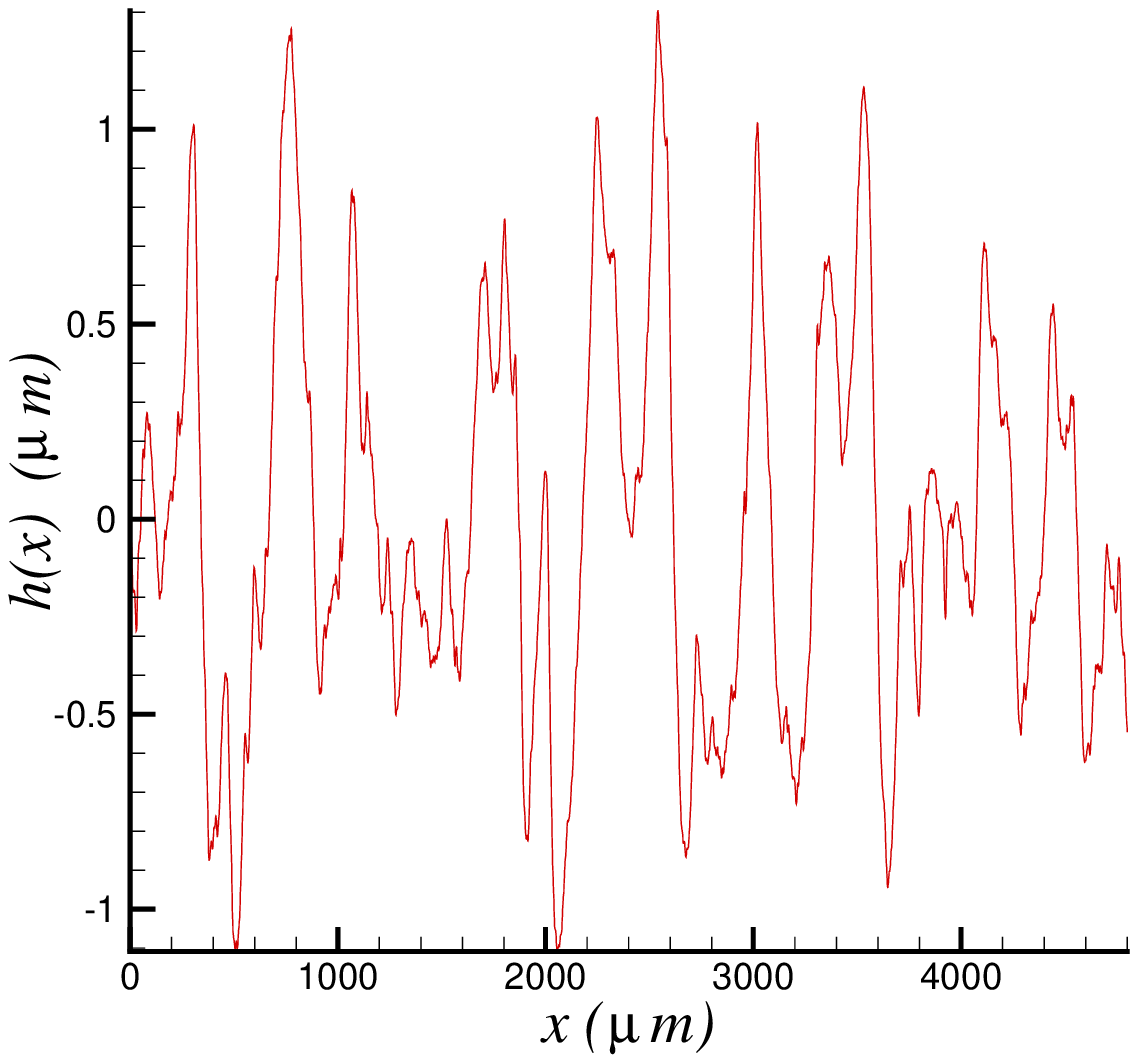}
\includegraphics[width=5.3cm,height=4.9cm,angle=0]{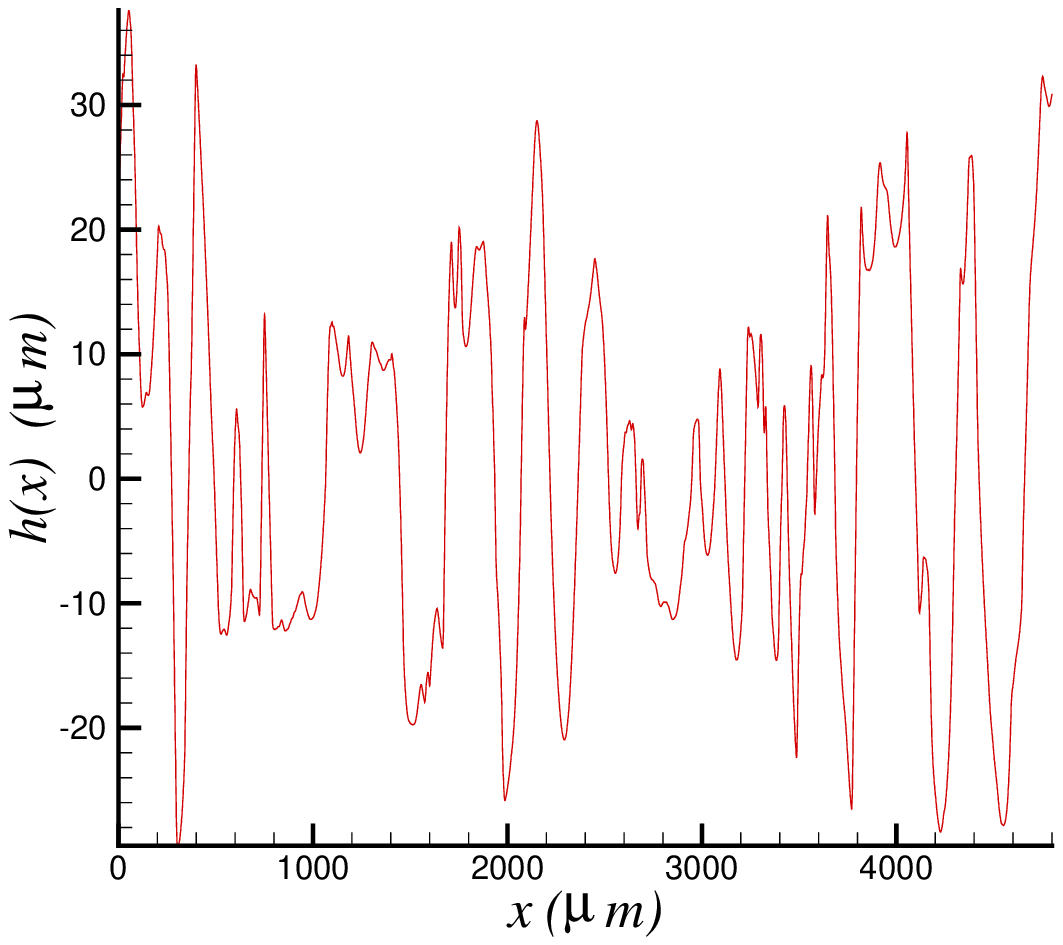}\hspace{4mm}(d) \caption{(a) and (b)
are AFM surface images in $5\times 5 \mu m^2 $ size of glass
films with etching time $2$ and $12$ minutes, respectively. (c) and (d) are Talysurf surface images
with $0.75 \mu m$, with etching times $20$ and $70$ minutes.}
\end{figure}
\endgroup
\end{center}
%%%%%%%%%%%%%%%%%%%%%%%%%%%%%%%%%%%%%%%%%%%%%%%%%%%%%%%%%%%%%%%%%%%%%%%%%%

\section{Experiments}

 We started with glass microscope slides as a sample. Only one side
of samples was etched by HF solution for various etching times, $2$ to $70$ minutes,
after cleaning by proper detergents. HF
concentration was $\%40$ for all experiments. The scattered light intensity of samples
was measured as a function of angle, $I(\theta)$ using a setup consisting of
a He-Ne laser ($\lambda=632.8nm$), a photo-multiplier tube (PMT) detector and
a computer controlled micro-stepper rotation stage. The resolution of
the micro-stepper was $0.5$ minutes per step. The surface topography of the
etched glass samples in small scale ($< 5 \sigma m)$ was obtained using an
atomic force microscope (AFM) (Park Scientific Instruments). The
images in small scale were collected in a constant force mode
and digitized into $256 \times 256$ pixels. A commercial standard pyramidal
$Si_3N_4$ tip was used. A variety of scans, each with size $L$, were
recorded at random locations on the surface. The large scale
($> 5 \mu m$) morphology line scans of the samples were recorded using
a surface profile-meter (Taylor Hobson). Fig. 1 shows typical
AFM image and surface profile data with resolutions of about $20 nm$
and $0.75 \mu m$, respectively. In order to directly characterize
different samples, we also used the Jascow spectrophotometer. In
Fig. 2 an illustration of the effect of the etching time on the
specular reflectivity measurement is presented. The reflectivity is
normalized to the smooth glass $R_0$, and represented as a function of
wavelength and etching time. As the etching time increases,
$\frac{R_S}{R_0}$ approaches to zero.
%%%%%%%%%%%%%%%%%%%%%%%%%%%%%%%%%%%%%%%%%%%%%%%%%%%%%%%%%%%%%%%%%%%%%%%%%%
\begin{figure}
\begin{center}
\epsfxsize=7truecm\epsfbox{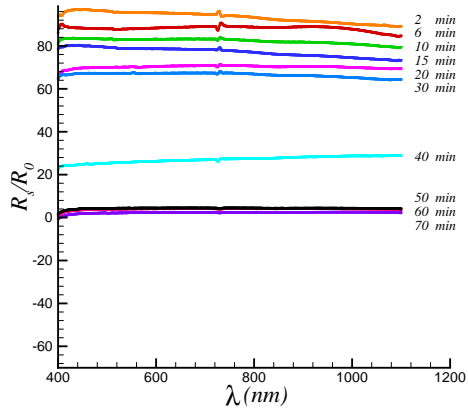}
 \caption{Specular reflection of glass surfaces after immersion in $40\%$
 HF for different time interval.}
\end{center}
 \end{figure}
%%%%%%%%%%%%%%%%%%%%%%%%%%%%%%%%%%%%%%%%%%%%%%%%%%%%%%%%%%%%%%
%%%%%%%%%%%%%%%%%%%%%%%%%%%%%%%%%%%%%%%%%%%%%%%%%%%%%%%%%%%%%%%%%%%%%%%%%%
\begin{figure}
\begin{center}
\epsfxsize=7truecm\epsfbox{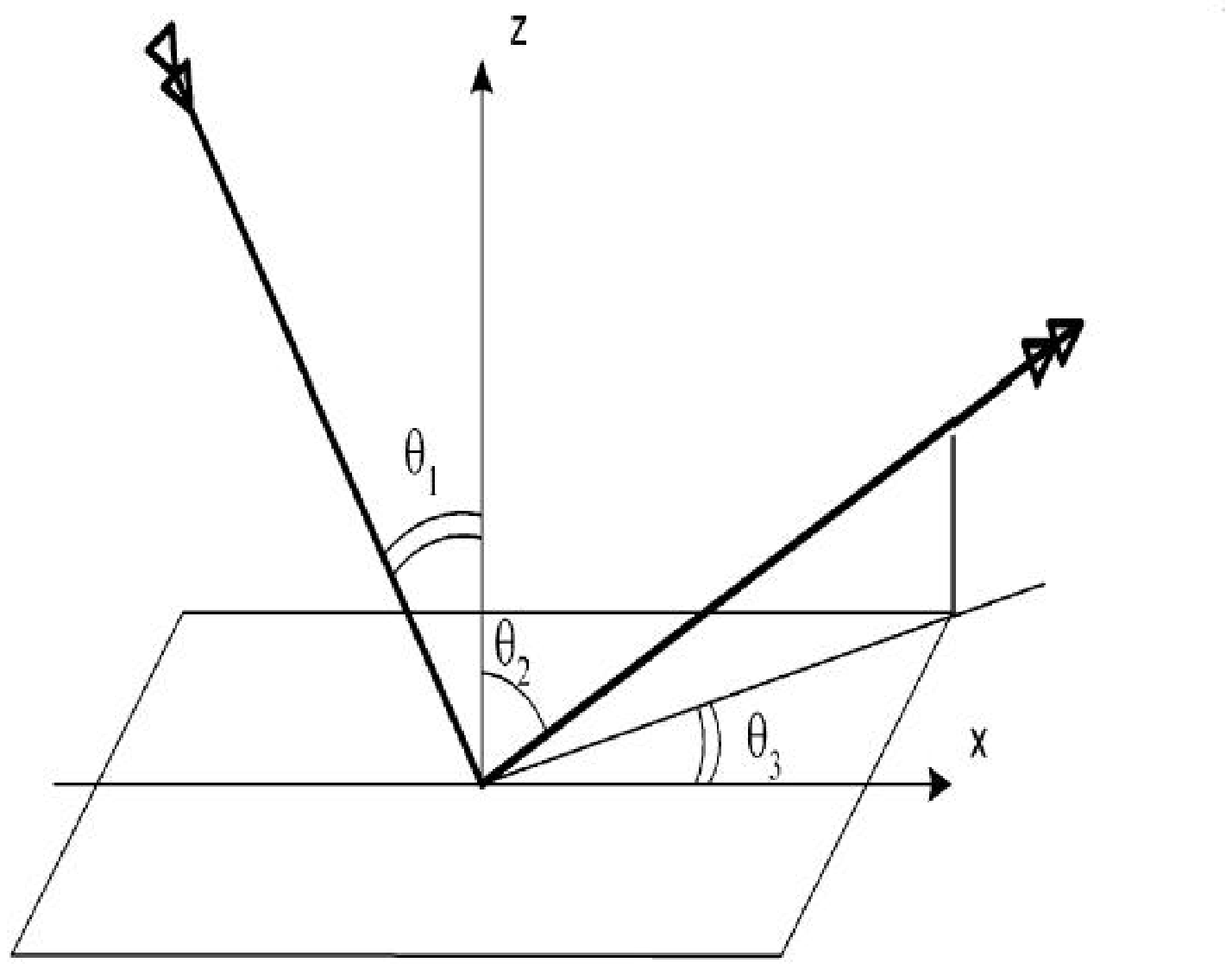}
 \caption{Scattering angles $\theta_1$, $\theta_2$ and $\theta_3$ for reflection.}
\end{center}
 \end{figure}
%%%%%%%%%%%%%%%%%%%%%%%%%%%%%%%%%%%%%%%%%%%%%%%%%%%%%%%%%%%%%%

\section{Theory}

\subsection{General notations}

 It is known that to derive the quantitative information of the
surface morphology one may consider a sample of size $L$ and define
the mean height $\overline{h}$ of the etching surfaces and
its roughness $\sigma$ by \cite{Marsilli}:
\begin{equation}\label{w}
\sigma(L,t) =(\langle (h-\overline{h})^2\rangle)^{1/2}
\end{equation}
Here $t$ is etching time, that is a factor which can apply to control
the glass surface roughness, and the $<\cdot \cdot \cdot>$ denotes an averaging
over different samples. In this work, we study the statistical
parameters of the surfaces. One of the possible initial conditions
is assuming to have a flat interface and the correlation function of
the rough surface is denoted by $C(R)$:
\begin{equation}\label{ising}
  C(R)=\frac{<h(r)h(r+R)>}{\sigma^{2}},
\end{equation}
and the correlation length $\lambda_0$ , is a distance at which
the correlation function falls by $1/e$.

\section{Kirchhoff's expressions for reflection}

Kirchhoff theory, also known as tangent plane or physical optics
theory is the most widely used in the study of wave scattering from
rough surfaces \cite{Ogilvy}. Kirchhoff theory has been applied to the study
of One and two dimensional exact approaches have been successfully
applied to dielectric, metallic or perfectly conducting surfaces
\cite{Kakuen, Nieto} dielectric films on a glass substrate \cite{Gu} and dielectric
films \cite{Ingve, Calvo}. Such exact calculations have been compared with
experimental results and approximate models \cite{Gu, Jerome}. Also some
authors studied wave scattering from random layers with rough
interfaces \cite{Antoine1, Antoine2}.

The Kirchhoff theory is based on three major assumptions: a) The
surface is observed from far field. b) The surface is regarded as
flat, and the optical behavior is locally identical to any given
point on the surface. Therefore the Fresnel laws can be locally
applied. c) The amplitude of the reflection coefficient, $R_0$ is
independent of the position on the rough surface. The field
scattered by the rough surface, $\psi_{sc}(r)$, considering far field
approximation is obtained by integration over the mean reference
plane $S_M$: (the geometry is displayed in Figure 3)
\begin{eqnarray}\label{3}
&& \psi^{sc}(r)=\frac{ik \exp(ikr)}{4\pi r}\int\int_{s_{M}}
(a\frac{\partial h}{\partial x_{0}}+b \frac{\partial h}{\partial
y_{0}}-c) \cr \nonumber \\
&& \exp{(ik(Ax_{0}+By_{0}+ C h(x_{0},y_{0})))} dx_{0}dy_{0}
\end{eqnarray}
where
\begin{eqnarray}\label{ABC}
A&=&\sin \theta_{1}-\sin \theta_{2} \cos \theta_{3} ,\cr \nonumber \\
B&=&-\sin \theta_{2} \sin \theta_{3} ,\cr \nonumber \\
C&=&-(\cos \theta_{1} +\cos \theta_{2}) ,\cr \nonumber \\
a&=&\sin \theta_{1}(1-R_{0})+\sin \theta_{2}\cos
\theta_{3}(1+R_{0}),\cr
\nonumber \\
b&=&\sin \theta_{2} sin \theta_{3}(1+R_{0}),\cr \nonumber \\
c&=&\cos \theta_{2}(1+R_{0})-\cos \theta_{1}(1-R_{0}) \cr
\nonumber
\end{eqnarray}

In the derivation of the Eq. (2), it is assumed that the
incident wave $\psi^{in}$ is a plane wave with a wave vector ${\bf
k}$ as $\psi^{in}(r)=\exp(i {{\bf k} } \cdot  {{\bf r}})$.
The height distribution function (PDF) and correlation function of
surfaces, which we used are Gaussian. It is shown that for such
surfaces the total scattered intensity can be written as \cite{Ogilvy}:
\begin{equation}
<I>= I_{0}\exp({-g}) + <I_{d}>
\end{equation}
Where $I_0$ and  are the scattered from a smooth surface and the
diffuse intensity respectively and $g$ can be written as: $g=k^{2}\sigma^{2}C^{2}$.
Therefore, we may divide them into three domains depends on the
value of $g$ as below:\\

\textbf{(a) Slightly rough surfaces:}\\

When the wave length is long enough compared to $\sigma$ or $g\ll 1$, the diffuse
 reflectance may be neglected. Therefore in specular direction the average intensity is \cite{Ogilvy}:
\begin{equation}
<I>\approx I_{0}\exp({-g}) + \frac{k^{2}F^{2}\lambda_{0}^{2}}{4\pi r^{2}}ge^{-g}A_{M}\exp(-\frac{k^{2}(A^{2}+B^{2})\lambda_{0}^{2}}{4})
\end{equation}
where $F$ is an angular factor given by: $F(\theta_1, \theta_2, \theta_3)=\frac{1}{2}(\frac{Aa}{C}+\frac{Bb}{C}+c)$\\

\textbf{b) Intermediate roughness surface}\\

Surfaces for which $g \sim 1$ are regarded as moderately rough. For this case,
it can be provided an upper and lower bound to the diffuse field intensity as \cite{Ogilvy}:
\begin{equation}
\frac{k^{2}F^{2}\lambda_{0}^{2}}{4\pi r^{2}}ge^{-g}A_{M}\exp(-\frac{k^{2}(A^{2}+B^{2})\lambda_{0}^{2}}{4})
< \langle I_0 \rangle \leq \frac{k^{2}F^{2}\lambda_{0}^{2}}{4\pi r^{2}}A_{M}
\end{equation}
where $A_M$ is the mean area of the scattering surface.\\

\textbf{c) Very rough surfaces}\\

To obtain a solution in this limit i.e. when $g\gg1$, we used the total scattered field
rather than the diffuse field\cite{Ogilvy}. For a Gaussian
correlation function, the diffuse field leads to:
\begin{equation}
\langle I \rangle\ \approx \frac{k^{2}F^{2}\lambda_{0}^{2}}{4\pi r^{2}}\frac{1}{g}
A_{M}\exp(-\frac{k^{2}(A^{2}+B^{2})\lambda_{0}^{2}}{4g})
\end{equation}
For very rough surfaces the coherent field will be negligible and Eq. (7) may be taken to
give the diffuse or total field from the scattering surfaces.
%%%%%%%%%%%%%%%%%%%%%%%%%%%%%%%%%%%%%%%%%%%%%%%%%%%%%%%%%%%%%%%%%%%%%%%%%%
\begin{figure}
\begin{center}
\epsfxsize=7truecm\epsfbox{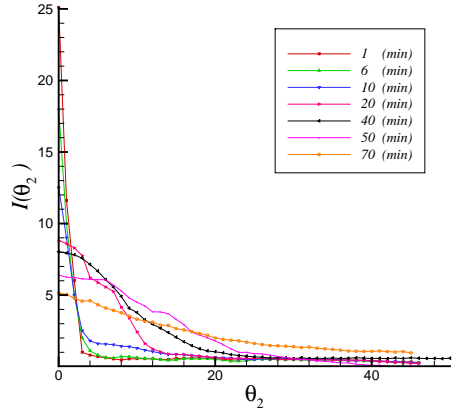}
 \caption{Some scattered normalized intensity distributions vs. scattering angles.}
\end{center}
 \end{figure}
%%%%%%%%%%%%%%%%%%%%%%%%%%%%%%%%%%%%%%%%%%%%%%%%%%%%%%%%%%%%%%

\section{Results and Discussion}

Referring to Fig. 2, it can be seen that as the etching time increases the specular
reflection of the samples are decreased. This depends on the surface roughness of the samples.
It has been shown that the $g$ parameter, Eq. (6), of the samples less than $12$ minutes
etching time are very less than unity, whereas for the samples with around $20$ minutes
etching time, $g$ is almost equal to one, and for those with $30$ and $40$ minutes etching
time $g$ is very larger than unity which are compatible with predicted domains by the
Kirchhoff theory. But the results of etched samples larger than about $50$ minutes,
cannot be explained by this theory. In this regime, geometric properties of light are more
dominant than its wave properties. The scattered intensity as a function of scattered
angle $\theta_2$ of some samples has been appeared on Fig. 4. It can be seen that, as it
is expected, the etching time of glass, affects on the intensity of scattered light.
To calculate the surface roughness, $\sigma$, and correlation length, $\lambda_0$, of the samples in the three
regimes we used the data from AFM and Talysurf. As it is known the resolution of the information from AFM is
between few nanometers to several micrometers, so for the samples with larger surface roughness
we had to use the data from Talysurf.
In Fig. 1, the illustration of the effect of etching time on the surface morphology
from AFM and Talysurf are presented. These figures exhibit an increasing of surface
roughness due to an increasing of etching time.
To compare the results from AFM and Talysurf with those calculated using the light scattering
for the three mentioned regimes, we list following results:

\textbf{a)} In the regime 1, for the samples with etching time less than $12$ minutes, Eq. (7)
was used to calculate $\sigma$.
 Fig. 5 shows the comparison of the surface roughness determined by AFM experiments and the one obtained by the
 Kirchhoff theory. According to this figure, the surface roughness is almost a linear function of etching time
of glass samples in a log-log coordinate for both data obtained from AFM and direct measurement of
scattered intensity. The agreement between both groups of data for samples with etching time less than $12$
 minutes is very good. At larger roughness, there are fairly disagreements between the theory
and experiment, because scattering plays a role in larger roughness and affects the measured
intensity in specular reflectance. If the surface roughness gets larger, the requirements of
the first domain of the Kirchhoff theory do not hold any more, then the roughness
from the reflection analysis and from AFM no
longer agree because of the increasing role of scattering.
%%%%%%%%%%%%%%%%%%%%%%%%%%%%%%%%%%%%%%%%%%%%%%%%%%%%%%%%%%%%%%%%%%%%%%%%%%
\begin{figure}
\begin{center}
\epsfxsize=7truecm\epsfbox{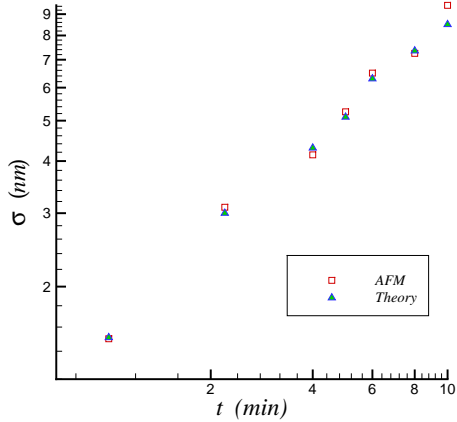}
 \caption{Comparison of the $\sigma$ surface-roughness (s) from AFM (square) and from reflection
 measurement (triangle) shows good agreement etching times less than or equal to $10$ minutes.}
\end{center}
 \end{figure}
%%%%%%%%%%%%%%%%%%%%%%%%%%%%%%%%%%%%%%%%%%%%%%%%%%%%%%%%%%%%%%

\textbf{b)} In the regime $2$, for the intermediate $g\approx 1$ and very rough surfaces $g\gg 1$, we are
dealing with the samples with etching time around $20$ minutes, and $30$-$50$ minutes,
respectively. Figuer 6 demonstrate $\sigma$ for different etching times of glass
that calculated by Talysurf data. To compare the given data with the results from the Kirchhoff
theory, both intermediate regime and very rough surface regime are compatible.
Fig. 7 shows the comparison between scattering spectrum obtained from the Kirchhoff theory of a very rough surface
 and that which is obtained from the experiment corresponding an etching time of $40$ minutes.
 %%%%%%%%%%%%%%%%%%%%%%%%%%%%%%%%%%%%%%%%%%%%%%%%%%%%%%%%%%%%%%%%%%%%%%%%%%
\begin{figure}
\begin{center}
\epsfxsize=7truecm\epsfbox{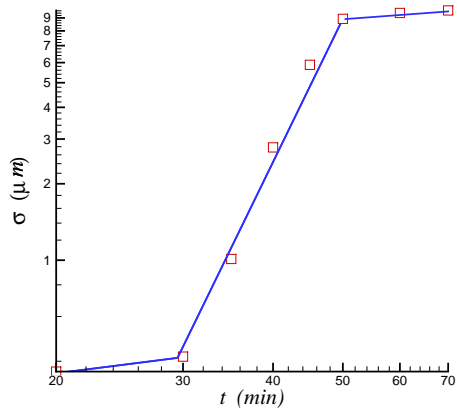}
 \caption{The variance of glass surfaces after
 immersion for different times by Talysurf.}
\end{center}
 \end{figure}
 %%%%%%%%%%%%%%%%%%%%%%%%%%%%%%%%%%%%%%%%%%%%%%%%%%%%%%%%%%%%%%%%%%%%%%%%%%
 \textbf{c)} In the regime $3$, it was found that for the samples with etching time larger than $50$ minutes the surface
 roughness and the correlation length are quite larger than the wave length for which the
 limitations of the Kirchhoff theory hold and it cannot be used. So, it can be considered that the
 rough surface is covered by some local smooth surfaces (i.e. meshes) with the size $(\sigma \times \lambda_0 )$,
 that these sizes are larger than the wave length. So, we used a theory based on geometric optics considerations.
 To describe the angular
 distribution of intensity of reflected light, we considered the angular distribution of the
 slope of rough surface, $p(\theta_2)$. To calculate $p(\theta_2)$, the data from Talysurf with resolution of
 $0.75 \mu m$ in horizontal steps were used.
 Fig. 8 presents comparison of normalized scattering field intensities obtained from the Kirchhoff
 theory and the one`obtained from an experiment with 70 minutes immersion.
 The agreement between geometric theory and experimental measurements
 is rather good.
%%%%%%%%%%%%%%%%%%%%%%%%%%%%%%%%%%%%%%%%%%%%%%%%%%%%%%%%%%%%%%%%%%%%%%%%%%
\begin{figure}
\begin{center}
\epsfxsize=7truecm\epsfbox{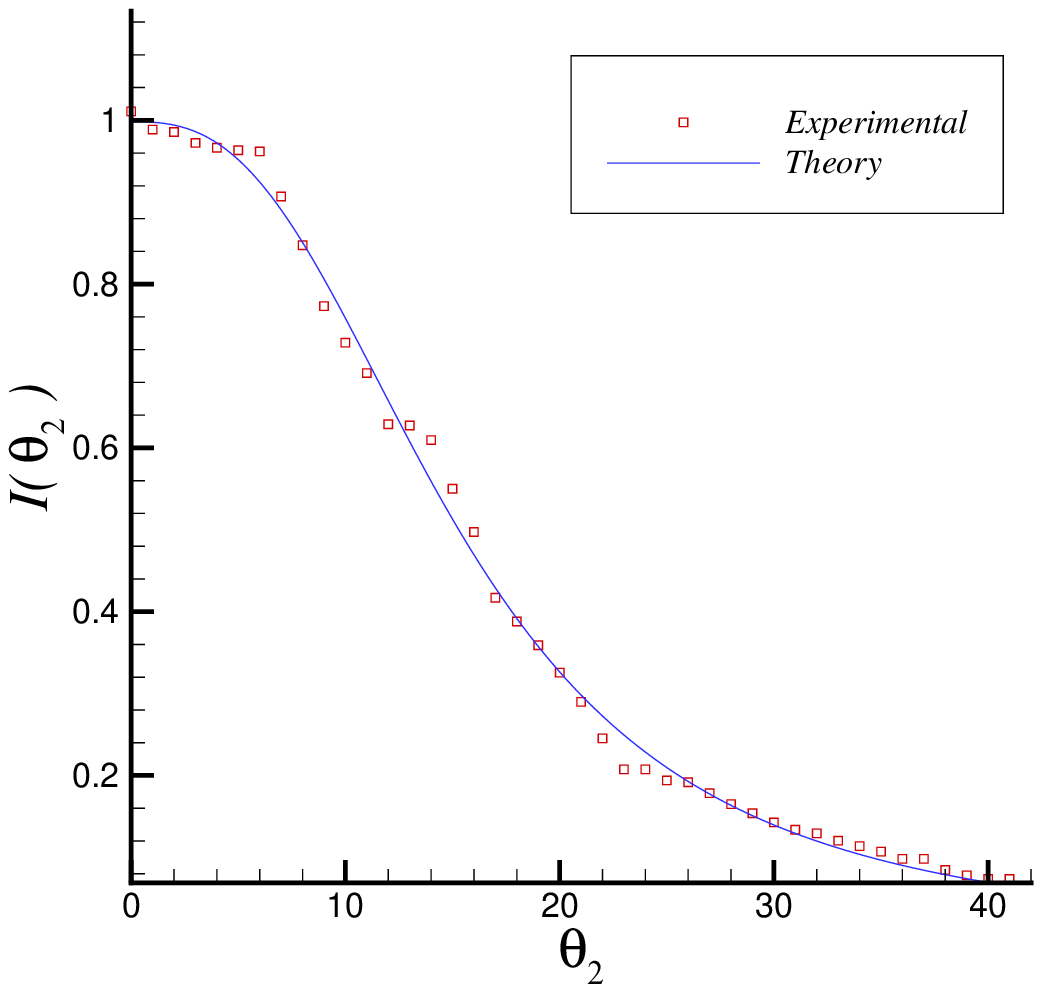}
 \caption{Comparison of measured diffuse field intensities with Kirchhoff theory and
 experiment after immersion for $40$ minutes.}
\end{center}
 \end{figure}
%%%%%%%%%%%%%%%%%%%%%%%%%%%%%%%%%%%%%%%%%%%%%%%%%%%%%%%%%%%%%%
%%%%%%%%%%%%%%%%%%%%%%%%%%%%%%%%%%%%%%%%%%%%%%%%%%%%%%%%%%%%%%%%%%%%%%%%%%
\begin{figure}
\begin{center}
\epsfxsize=7truecm\epsfbox{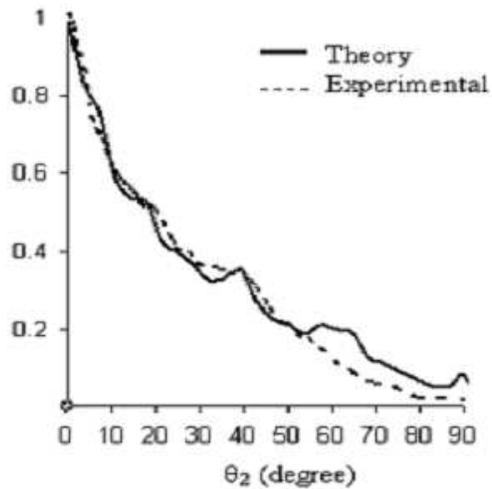}
 \caption{Comparison of normalized scattering field intensities obtained from the Kirchhoff
 theory and the one`obtained from an experiment with 70 minutes immersion.}
\end{center}
 \end{figure}
%%%%%%%%%%%%%%%%%%%%%%%%%%%%%%%%%%%%%%%%%%%%%%%%%%%%%%%%%%%%%%
The comparison between geometric reflection theory and experiment is shown on Fig. 11.

%\section{Conclusion}

Summarizingly, we studied the glass etching process by the optical scattering and their statistical
properties. By increasing the etching time for a series of the samples, we have studied optical
scattering, identified by reflection measurements with a spectrophotometer and a setup discussed
in experimental section. We found that these properties are explained by three regimes during etching.
In the other view, statistical properties of their surfaces which are found by the AFM and Talysurf,
confirm with these regimes. The roughness, which is obtained by the optical scattering, has a good agreement
with the experimental results.

\section{acknowledgment}

we would like to thank M. Shirazi for her useful comments and
discussions.
%$$$$$$$$$$$$$$$$$$$$$$$$$$$$$$$$$$$$$$$$$$$$$$$$$$$$$$$$$$$
%$$$$$$$$$$$$$$$$$$$$$$$$$$$$$$$$$$$$$$$$$$$$$$$$$$$$$$$$$$$

\end{document}